\documentclass[a4paper,twoside]{article}
\usepackage{AMS2LA,fancyhdr,CJK,multicol,graphicx,bm,ccmap,float}
\usepackage[dvipdfm, a4paper, pdfstartview=FitH, bookmarks=false, colorlinks=false,
        linkcolor=blue, urlcolor=blue, pdfborder=001, citecolor=blue,
        pdftitle={Changepdftitle}, pdfauthor={Changepdfauthor},
        pdfcreator=CHIN.\ PHYS.\ LETT., pdfproducer=CPL,
        pdfkeywords={changePACS}]{hyperref}
\def\headrule{\kern 1mm \hrule width 17cm \kern -1mm}%
\def\footnoterule{\kern 1mm \hrule width 7cm \kern 2.2mm}%
\def\REF#1{\par\hangindent\parindent\indent\llap{#1\enspace}\ignorespaces}%

\newcommand{\cplyear}{2015} \newcommand{\cplvol}{XX}
\newcommand{\cplno}{X} \newcommand{\cplpagenumber}{XX{XXXX}}
 \newcommand{\cplpage}{\cplpagenumber-\thepage}
\def\celsius{\ensuremath{^\circ\hspace{-0.09em}\mathrm{C}}}

\begin{document} \begin{CJK}{UTF8}{gbsn}\vspace* {-6mm} \begin{center}
\large\bf{\boldmath{Temperature Effects on Information Capacity and Energy Efficiency of Hodgkin-Huxley Neuron}}
\footnote{This work was supported by the National Natural Science Foundation
of China under Grant Nos. 11105062, 11275003, 11265014 and 11275084, the Fundamental Research Funds for the Central Universities under Grant No. lzujbky-2015-119.

\hspace*{1.8mm}$^{**}$Correspondence author. Email:
yulch@lzu.edu.cn

\hspace*{1.8mm}\copyright\,{\cplyear}
\href{http://www.cps-net.org.cn}{Chinese Physical Society} and
\href{http://www.iop.org}{IOP Publishing Ltd}}
\\[4mm]
\normalsize \rm{}Wang Long-Fei(王龙飞)$^{1}$, Jia Fei(贾斐)$^{2}$, Liu Xiao-Zhi(刘效治)$^{2}$, Song Ya-lei(宋亚磊)$^{3}$,\\ Yu Lian-Chun(俞连春)$^{1**}$
\\[1mm]\small\sl
$^{1}$Institute of Theoretical Physics, Lanzhou University, Lanzhou 730000, China

$^{2}$Cuiying Honors College, Lanzhou University, Lanzhou 730000, China

$^{3}$Institute of Computational Physics and Complex Systems, Lanzhou University, Lanzhou 730000, China
\\[4mm]\normalsize\rm{}(Received xx July 2015)
\end{center}
\end{CJK}
\vskip -1mm

\noindent{\narrower\small\sl{}
Recent experimental and theoretical studies show that energy efficiency, which measures the amount of information processed by a neuron with per unit of energy consumption, plays an important role in the evolution of neural systems. Here, we calculated the information rates and energy efficiencies of the Hodgkin-Huxley (HH) neuron model at different temperatures in a noisy environment. We found that both the information rate and energy efficiency are maximized by certain temperatures. Though the information rate and energy efficiency cannot be maximized simultaneously, the neuron holds a high information processing capacity at the temperature corresponding to maximal energy efficiency. Our results support the idea that the energy efficiency is a selective pressure that influences the evolution of nervous systems.
\par}\vskip 3mm\normalsize

\noindent{\narrower\sl{PACS:87.19.ly, 87.19.ls, 87.19.lc, 87.16.Vy}
{\rm\hspace*{13mm}DOI:10.1088/0256-307X/\cplvol/\cplno/\cplpagenumber}

\par}\vskip 3mm

\begin{multicols}{2}

Information processing in the nervous system is metabolically expensive. Human brain is only about 2\% of the total body weight, but consumes about 20\% of the resting metabolic energy.$^{[1]}$ The large metabolic energy requirement of nervous system could constrains the size and structure of the brain, and may have largely optimized the nervous system by favouring energy efficient neural morphologies, codes, wiring patterns and brain structures.$^{[2,3]}$ The energy efficiency of nervous system have possibly been greatly optimized by natural selection.

There are many factors that influence the energy efficiency. Of all the information processing activities, action potential (AP) makes a great contribution of total consumed energy.$^{[1,4]}$ AP on its own does not require energy, however, restoring the transmembrane ionic concentration gradients through the $Na^+$/$K^+$ ion pump is an energy consuming process which relies on the energy released by adenosine triphosphate (ATP) hydrolysis. In the classic Hogkin-Huxley (HH) model, due to overlap between $Na^+$ entry and $K^+$ outflow at the same time, the flow of $Na^+$ and $K^+$ ions during AP largely exceed the minimum required by a pure capacitor, and thus waste large amounts of energy to restore the membrane potential.$^{[5-7]}$ Recent investigations on the nonmyelinated mossy fibers of the rat hippocampus shows that minimizing the overlap of these two ion fluxes improves the energy efficiency of AP generation.$^{[8]}$
Energy efficiencies of the neural systems are also constrained by their size. Recent theoretical and numerical studies suggest that in the noisy environment the energy efficiency is maximized by the number of ion channels in a single neuron, or the number of neurons in a neuronal circuits$.^{[9,10]}$

Temperature is an important factor that constrain the energy efficiency of neural systems. It influences the conductance, activation and inactivation of all ion channels as a global perturbation.$^{[11]}$ Temperature sensitivities of ion channels present a challenge to maintaining stable functions over an extended temperature range$.^{[12]}$ Yu \textit{et al.}, found that warmer body temperatures reduce the $Na^+$/$K^+$ overlap and facilitate single energy efficient action potentials.$^{[13]}$

However, information in neural system is often represented in groups of APs generated by neuron population or a single neuron within a time period, rather than a single AP. Therefore, it is interesting to study the effects of temperature on the energy efficiency in processing input signals with the consideration of noise perturbation. In this research, we stimulate single compartment HH neuron with synaptic inputs in presence of Gaussian white noise. The effective energy efficiency rate was introduce to measure how much information is encoded in a unit time using unit energy in response to the synaptic stimuli. We then investigated the influence of temperature on the energy efficiency of information processing by the HH neuron and found that the energy efficiencies exhibit a maximum at certain temperatures.

\begin{figure*}
\begin{center}
     \includegraphics[width=1.0\textwidth]{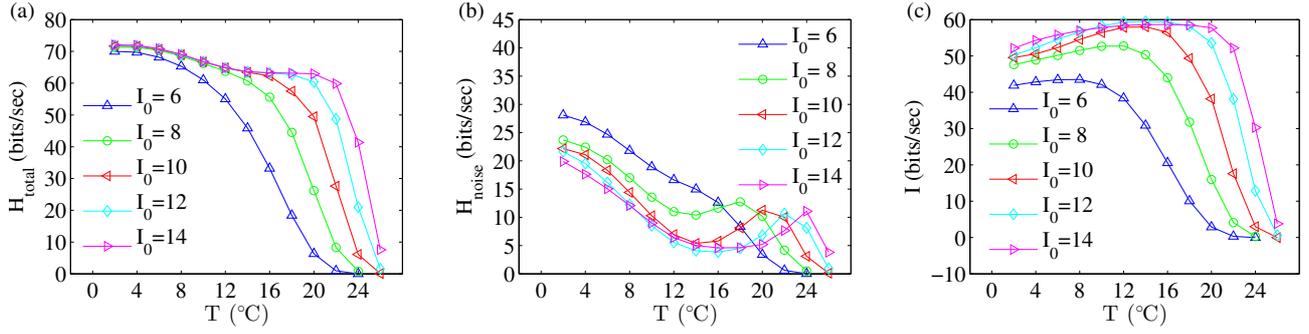}
\end{center}
\caption{The information capacity of the HH neuron as a function of temperature for different input signal strength (unit: ${\rm \mu A/cm^2}$). (a) total entropy rate, (b) noise entropy rate, and (c) the information rate.}
\label{H}
\end{figure*}

The single compartment HH neuron with Gaussian white noise $\xi(t)$ and external current injection $I_e(t)$ is described by following equations.$^{[14]}$
\begin{eqnarray}
C_m\frac{d V}{d t} & = & -\bar{g}_{Na}m^{3}h(V-V_{Na}) -\bar{g}_{K}n^{4}(V-V_{K})\nonumber\\
 	&  & -\bar{g}_{L}(V-V_{L})+\xi(t) +I_e(t),\label{hh_v}\\
\frac{d x}{d t} & = & \alpha_x(V)(1-x)-\beta_x(V) \,\, (x=m,h,n).\label{hh_mhn}
\end{eqnarray}
Eq.\,(1) shows the varations in membrane potential ($V$) over time where $C_m=1{\rm \mu F /cm^2}$ is the membrane capacitance. The maximum channel conductances and reversal potentials are typical values: $\bar{g}_{Na}=12{\rm 0mS/cm^2}$, $\bar{g}_K=36.0{\rm mS/cm^2}$, $\bar{g}_L=0.33{\rm mS/cm^2}$,k  $V_{Na}=50{\rm mV}$, $V_K=-77{\rm mV}$ and $V_L=-54.4{\rm mV}$. $m^3h$ and $h^4$ are ratios of sodium channels and potassium channels at open state to all sodium channels and potassium channels, respectively. $\xi (t) $ is the zero-mean Gaussian white noise with
 \begin{equation}
<\xi(t)>=0; <\xi(t)\xi(t')>=2D\delta (t-t'),
\end{equation}
where $D$ is the noise intensity. $I_e(t)=\sum_j I^j_{syn}(t)$ is composed of many single synaptic-current-like pulses $I^j_{syn}(t)$ , a single pulse have the form
\begin{equation}
I^j_{syn}(t) =
\left\{
  \begin{array}{ll}
   I_0 (t-t^j_s)\mathrm{e}^{\frac{t-t^j_s}{\tau}}, &  t^j_s \leq t \leq t^j_c,\\
   0, &  t<t^j_s \,\,or\,\, t>t^j_c,
   \end{array}
\right.
\end{equation}
where $j$ denote the $j$th pulse, $I_0$ represents the strength of the synaptic-current-like pulse, $t^j_s$, $t^j_c=t^j_s+\Delta t$ ($\Delta t=8{\rm ms}$ in our simulation) are the time when released neural transimitters first contact with the post-synaptic membrane (the pulse start) and the cut-off time of the pulse (the pulse end), representatively. $\tau=2{\rm ms}$ is the time delay constant. The spiking time of pulses follow Poisson distribution with the average time interval $100{\rm ms}$ as expected for a real neuron.$^{[15]}$

Eq.\,(2) represents three similar gating equations of $Na^+$ and $K^+$ channels, $x$ can be any activation/inactivation variables $m$, $h$ or $n$. The voltage-dependent rate functions of gating states transition $\alpha_x(V)$ and $\beta_x(V)$ are
\begin{eqnarray}
&\alpha_{m} = \phi(T)0.1\frac{25-V}{\mathrm{e}^{(25-V)/10}-1}, & \beta_{m}  = \phi(T)4.0\mathrm{e}^{-\frac{V}{18}}; \nonumber\\
&\alpha_{n} = \phi(T)0.01\frac{10-V}{\mathrm{e}^{(10-V)/10}-1}, & \beta_{n} = \phi(T)0.125\mathrm{e}^{-\frac{V}{80}};\nonumber\\
&\alpha_{h} = \phi(T)0.07\mathrm{e}^{-\frac{V}{20}}, & \beta_{h} = \phi(T)\frac{1}{\mathrm{e}^{\frac{30-V}{10}}+1}. \nonumber
\end{eqnarray}
where $\phi (T)$ is the temperature dependent function$^{[13]}$
\begin{equation}
\phi(T)=3.0^{(T-6.3)/10}.
\end{equation}

The equations are numerically integrated using Euler method with time step $dt=0.01ms$. The total simulation time lasted for $3600{\rm s}$ in order to estimate the information entropy (see below).

\begin{figure*} 
\begin{center}
      \includegraphics[width=1.0\textwidth]{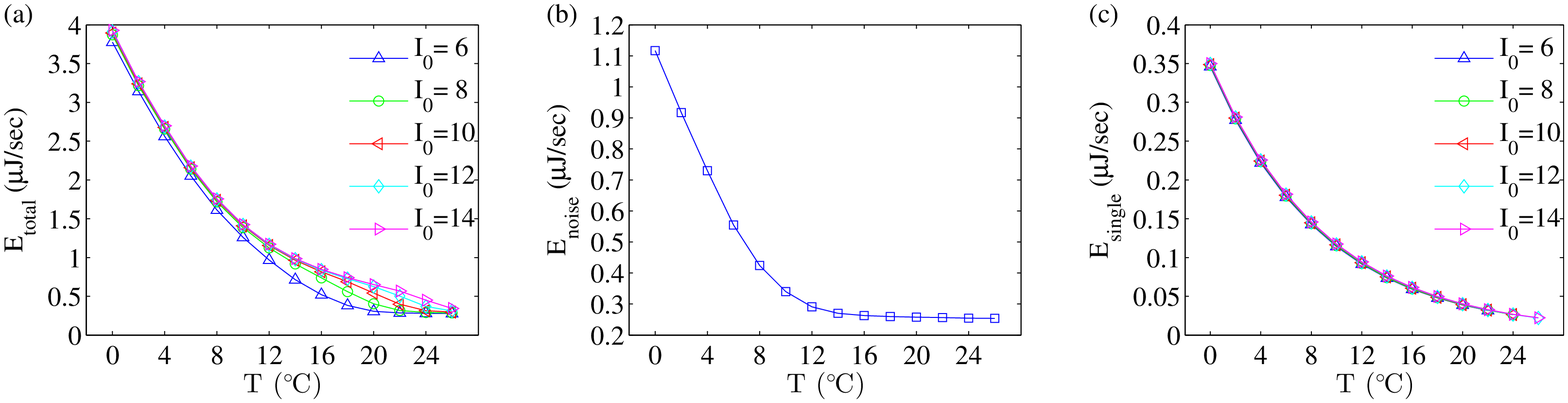}
\end{center}
\caption{Energy consumption rates as a function of temperatures of the HH neuron. (a) total energy consumption rates; (b) energy consumption rates induced by noise alone; (c) the energy cost of a single AP. The legend of A and C denote the stimus strength with unit ${\rm \mu A/cm^2}$.}
\label{e}
\end{figure*}

To measure the average information processing efficiency of our HH neuron under the mimicked external synaptic pulses, we calculated the information entropy rate and information rate using the method of Strong {\it et al}.$^{[16]}$

We first calculated the entropy rate of the spike train. First, we divided the spike train in response to set stimulus into bins of size $\Delta\tau$ with value 1 (AP) or 0 (no AP). Then we divided the bin train into $T/\Delta\tau$ words. Considering the size of data set, the total entropy rate is defined as
\begin{equation}
H_{total}=-T\lim_{T\rightarrow\infty}\frac{1}{T}\sum_iP_i\mathrm{log}_2P_i,
\end{equation}
where $P_i$ represents the probability of $i$-th word. Similar to the total entropy rate, we evaluate the information rate interrupted by noise using average noise entropy rate
\begin{equation}
H_{noise}=<-\lim_{T\rightarrow\infty}\frac{1}{T} \sum_i P_i(t)\mathrm{log}_2P_i(t)>_t,
\end{equation}
where $P_i(t)$ is the time-dependent word probability. We then calculated information rate, which measures the average information encoded into the spike trains in response to stimulus, by subtracting noise entropy rate from total entropy rate.
\begin{equation}
I=H_{total}-H_{noise}.
\end{equation}

Fig.\,1 (a) , (b) and (c) demonstrate the total entropy rate, noise entropy rate, and information rate as functions of temperature for HH neuron in response to signals with different strength, respectively. We observed that both the total entropy rate and noise entropy rate decrease as the temperature increases. However, when temperature is high, the total entropy rate decreases rapidly, whereas the noise entropy rate shows a local maxima. The information rate, i.e. the difference between total entropy rate and noise entropy rate, exhibits a global maxima at certain temperatures between 0 to 25$\celsius$.

As shown in Fig.\,1, for strong signals, the total entropy rate is high but the noise entropy is low. This may be because strong signals can induce more APs, which in turn increases the total information received by the neuron (Fig.\,1 (a)). Strong signals also have high probability to induce APs, which increases the coherence of firings across the trials, thus decrease the noise entropy in the spike trains (Fig.\,1 (b)). As a result, the information rate is high for strong signals and decreases as signal strength decrease (Fig.\,1 (c)).

We also calculated the energy consumption rate of the HH neuron in the above signal processing event. The energy cost of a neuron are calculated mostly by integrating the $Na^+$ or $K^+$ current over time, and then convert them into the number of $Na^+$ or $K^+$ ions. Since the $Na^+$/$K^+$ pump hydrolyzes one ATP molecule for three $Na^+$ ions extruded and two $K^+$ ions imported, the energy cost is estimated by the amount of ATP molecules are expended, or in units of energy if the powers of the $Na^+$/$K^+$ pump are measured in real experiment (for example, $50 {\rm kJ/mol}$ in the heart ).$^{[13]}$ Recently, Moujahid {\it et al.} proposed a novel method to estimate the energy cost directly from the equivalent electrical circuit of the HH neuron.$^{[17,18]}$ Here in this paper we use a similar but more direct method to calculate the energy cost of the above HH neuron in processing signals. We consider the energy provided by or leak out to external environment. In the equivalent electrical circuit of HH neuron, the energy is consumed by three types of ion channel conductance and can be provided or consumed by external electrode-clamping circuits. Thus the total energy consumption rate is equal to the sum of all four powers
\begin{eqnarray}
\frac{dE(t)}{dt} & = & p_e(t)+p_{Na}(t)+p_K(t)+p_L(t)\\
                 & = & VI_e(t) +i_{Na}(V-V_{Na})\nonumber \\
                 &   & +i_K(V-V_K)+i_L(V-V_L)\\
                 & = & VI_e(t)-\bar{g}_{Na}m^{3}h(V-V_{Na})^2 \nonumber \\
                 &   & -\bar{g}_{K}n^{4}(V-V_{K})^2-\bar{g}_{L}(V-V_{L})^2. \label{energy}
\end{eqnarray}

In Eq.\,(11), the minus sign in front of ion channel terms represent energy consumption while ignored plus sign in front of external stimuli term represents power supply. To calculate total energy consumption, we reversed the signs on the right hand side of Eq.\,(11) to denote energy consumption of ion channels as positive and power supplied (or consumption if the value of $VI_e(t)$ itself is negative) by external circuits as negative. To estimate $E_{total}$, the total energy cost per unit time in the signal detection progress, $dE(t)/dt$ is integrated over time and then divided by total simulation time. The energy cost rate induced by noise alone ($E_{noise}$) is calculated as above but without input signals. The energy cost of a single AP ($E_{single}$) is calculated by integrating $dE(t)/dt$ in the period of the AP.

As demonstrated in Fig.\,2 (a) , the total energy consumption rate decreases as temperature increases. At lower temperature, the energy consumption rate is almost identical for different signal strengths. At high temperature, the neuron consumes more energy to processing strong signals. The dependence of $E_{noise}$ on temperature decreases with temperature and has two distinct stages: at lower temperature (less than $\rm 15\celsius$) it decreases almost lineally and at higher temperature (above $\rm 15\celsius$) it almost flattens (Fig.\,2 (b)). As displayed in Fig.\,2(c), the average energy cost of single AP is identical to different signal strength, because the shape of AP almost does not change as signal changes. As the temperature increases, the energy cost of an AP decreases, which is in line with previous studies. $^{[10,13]}$

\begin{figure}[H]
\begin{center}
	\includegraphics[width=0.49\textwidth]{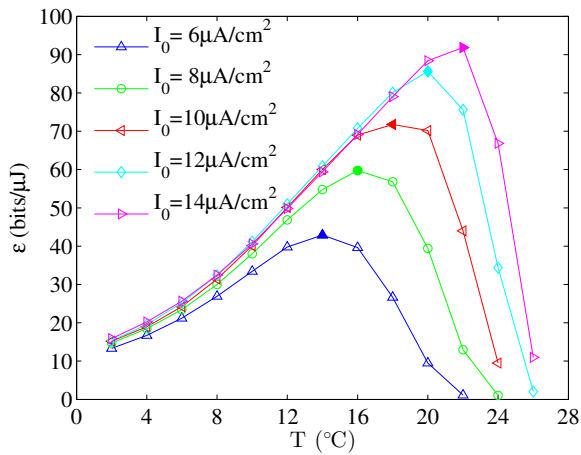}
\end{center}
\caption{Energy efficient as a function of temperature for different input signal strengths. The solid markers denote the maximum energy efficiency.}
\label{ee}
\end{figure}
Finally, we define the energy efficiency as the information rate divided by energy consumption rate, i.e.
\begin{equation}
\varepsilon =\frac{I}{E},
\end{equation}
which measures how much information is effectively processed by the neuron in response to the stimulus and consuming one unit of energy. As shown in Fig.\,3, with the increase in temperature, the energy efficiency first increases and then decreases, passing through a global maximum as a function of temperature for different signal strength. The existence of maximum implies that the temperature could maximize the energy efficiency of the neuron in the processing of input signals. Meanwhile, with the increasing signal strength, the energy efficiency increases, as well as the temperature corresponding to maximal energy efficiency.

Comparing data at the optimal temperature, i.e. the temperature with maximum energy efficiency, and at the temperature with maximum information rate (see Table.\,1, variables at the optimal temperature are marked with superscript `o'), we found that the temperatures that maximize energy efficiency ($T^{o}$) and that which maximize information rate ($T^{I}$) for a particular signal strength is not the same; information rate and energy efficiency cannot be maximized by the temperature simultaneously. However, at the temperature the energy efficiency is maximized, the corresponding information rate is in the range of $70\%$ to $90\%$ of the maximal information rate ($\frac{I^o}{I^{max}}$). The signal detection rate $P^o_{detection}$, which is ratio of the number of APs induced immediately (less thant $8 ms$) after signals are applied and the total number of applied signals at $T^{o}$, is in the range from $80\%$ to $99\%$. Thus, it is concluded that at the temperature the energy efficiency is maximized, the neuron can still keep relatively high capacity to process information. At the same time, at the optimal temperature $T^{o}$, the spontaneous firing rate is very low ($f^o_{noise} < 0.27 Hz $ ), and the noise induced energy cost is less than $40\%$ of total energy cost ($\frac{E^o_{noise}}{E^o}$).

\noindent{\footnotesize Table 1. Variables at optimal temperature

\vskip 2mm \tabcolsep 2pt

\centerline{\footnotesize
\begin{tabular}{ccccccc}\hline
I(${\rm \frac{\mu A}{cm^2}}$) & $T^o$($\celsius$) & $T^{I}$($\celsius$) & $\frac{I^o}{I^{max}}$ & $P^o_{detection}$ & $f^o_{noise}$(Hz) & $\frac{E^o_{noise}}{E^o_{total}}$\\
\hline
6 & 14 & 6 & 0.7094 & 0.7922 & 0.262 & 0.3012 \\
8 & 16 & 10 & 0.8341 & 0.9367 & 0.075 & 0.2943 \\
10 & 18 & 12 & 0.8509 & 0.9766 & 0.018 & 0.3219\\
12 & 20 & 12 & 0.8998 & 0.9894 & 0.005 & 0.3573\\
14 & 22 & 14 & 0.8898 & 0.9941 & 0 & 0.3954\\
\hline
\end{tabular}
}
}

\vskip 0.5\baselineskip

In conclusion, we investigated the influence of temperature on the capacity of information processing and energy efficiency of the classic HH neuron in response to synaptic inputs. We found that temperature can maximize the information capacity, as well as the energy efficiency of HH neuoron. The difference between the temperature that maximizes energy efficiency and that which maximizes information rate for a particular signal strength produces a conflict; information rate and energy efficiency cannot be maximized simultaneously. This conflict has been observed in models of spiking neurons and neural codes,$^{[10, 19, 20]}$ adding to the numerous lines of evidence suggesting that energy is a selective pressure that has influenced the evolution of neural systems.$^{[21]}$ Our study further show that at the temperature the energy efficiency is maximized, the neuron can still keep its information processing capacity high, comparing its values at maximum. Considering that the typical living temperature of {\it Loligo} (the squid used by Hogkin and Huxley$^{[14]}$) is 10-26$\celsius$$^{[22,23]}$, our result of optimized temperature is consistant with real data.

\section*{\Large\bf References}

\vspace*{-0.8\baselineskip}

\hskip 7pt {\footnotesize

\REF{[1]} Attwell D and Laughlin S B 2001 {\it J. Cereb. Blood Flow Meta.} {\bf 21} 1133

\REF{[2]} Niven J E and Laughlin S B 2008 {\it J. Exp. Biol.} {\bf 211} 1792

\REF{[3]} Laughlin S B 2001 {\it Curr. Opin. Neurobiol.} {\bf 11} 475

\REF{[4]} Howarth C, Gleeson P and Attwell D 2012 {\it J. Cereb. Blood Flow Meta.} {\bf 32} 1222

\REF{[5]} Sengupta B, Stemmler M, Laughlin S B and Niven J E 2010 {\it PLoS Comput. Biol.} {\bf 6} e1000840

\REF{[6]} Hodgkin A L 1975 {\it Phil. Trans. R. Soc. B} {\bf 270} 297

\REF{[7]} Crotty P, Sangrey T and Levy W B 2006 {\it J. Neurophysiol. } {\bf 96} 1237

\REF{[8]} Alle H, Roth A and Geiger J R P, 2009 {\it Science} {\bf 325} 1405

\REF{[9]} Schreiber S, Machens C K, Herz A V M and Laughlin S B 2002 {\it Neural Comput.} {\bf 14} 1323

\REF{[10]} Yu L and Liu L 2014 {\it Phys. Rev. E.} {\bf 89} 032725

\REF{[11]} Hille B 2001 {\it Ion channels of excitable membranes} (Sunderland, MA: Sinauer)

\REF{[12]} Rinberg A, Taylor A L and Marder E. 2013 {\it PLoS Comput. Biol.} {\bf 9(1)} e1002857

\REF{[13]} Yu Y, Hill A P and McCormick D A 2012 {\it PLoS Comput. Biol. } {\bf 8} e1002456

\REF{[14]} Hodgkin A L and Huxley A F 1952 {\it J. Physiol.} {\bf 117} 500

\REF{[15]} Sengupta B, Faisal A A , Laughlin S B and Niven J E 2013 \emph{J. Cereb. Blood Flow Meta.}, \textbf{33}, 1465

\REF{[16]} Strong S P, Koberle R, van Steveninck R R R and W Bialek 1998 {\it Phys. Rev. Lett.} {\bf 80} 197

\REF{[17]} Moujahid A, d'Anjou A, Torrealdea F J and Torrealdea F 2011 {\it Phys. Rev. E } {\bf 83} 031912

\REF{[18]} Hasegawa H 2011 arXiv:1106.5862[cond-mat.dis-nn]

\REF{[19]} Balasubramanian V, Kimber D and Berry M J 2001 {\it Neural Comput.} \textbf{13} 799

\REF{[20]} De Polavieja G G, 2002 {\it J. Theor. Biol.} {\bf 214} 657

\REF{[21]} Niven J E and Farris S M 2012 {\it Curr. Biol.} {\bf 22} R323

\REF{[22]} Jacobson L 2005 {\it NOAA Technical memorandum} {\bf NMFS-NE} 193

\REF{[23]} McMahon J J and Summers W C 1971 {\it Biol. Bull.} {\bf 141} 561
}

\end{multicols}
\end{document}